\documentclass[aip,apl,floatfix,reprint,groupedaddress]{revtex4-1}
\usepackage{graphicx}
\usepackage{epstopdf}
\usepackage{epsfig}
\usepackage{amsmath}

\begin{document}


\title{Electric-field-driven domain wall dynamics in perpendicularly magnetized multilayers} 



\author{Diego L\'opez Gonz\'alez,$^{1}$ Yasuhiro Shirahata,$^{2}$ Ben Van de Wiele,$^{3}$ K\'{e}vin J. A. Franke,$^{1}$ Arianna Casiraghi,$^{1}$  Tomoyasu Taniyama,$^{2}$ and Sebastiaan van Dijken$^{1}$}
\email[]{sebastiaan.van.dijken@aalto.fi}
\affiliation{$^{1}$NanoSpin, Department of Applied Physics, Aalto University School of Science, P.O. Box 15100, FI-00076 Aalto, Finland.}
\affiliation{$^{2}$Materials and Structures Laboratory, Tokyo Institute of Technology, Yokohama, Japan.}
\affiliation{$^{3}$Department of Electrical Energy, Systems and Automation, Ghent University, Ghent B-9000, Belgium.}


\date{\today}

\begin{abstract}
We report on reversible electric-field-driven magnetic domain wall motion in a Cu/Ni multilayer on a ferroelectric BaTiO$_3$ substrate. In our heterostructure, strain-coupling to ferroelastic domains with in-plane and perpendicular polarization in the BaTiO$_3$ substrate causes the formation of domains with perpendicular and in-plane magnetic anisotropy, respectively, in the Cu/Ni multilayer. Walls that separate magnetic domains are elastically pinned onto ferroelectric domain walls. Using magneto-optical Kerr effect microscopy, we demonstrate that out-of-plane electric field pulses across the BaTiO$_3$ substrate move the magnetic and ferroelectric domain walls in unison. Our experiments indicate an exponential increase of domain wall velocity with electric field strength and opposite domain wall motion for positive and negative field pulses. Magnetic fields do not affect the velocity of magnetic domain walls, but independently tailor their internal spin structure, causing a change in domain wall dynamics at high velocities. 
\end{abstract}


\maketitle 

Controlled motion of domain walls in perpendicularly magnetized layers forms the basis of spintronic memory and logic device concepts.\cite{CUR-12,PAR-15} The technological relevance of materials with perpendicular magnetic anisotropy (PMA) stems from the stability and small width of their domain walls. Domain walls in PMA nanowires are efficiently driven by electrical currents via spin-orbit induced effects.\cite{MIR-11,EMO-13,RYU-13} Motivated by the prospect of low-power consumption and the ability to tailor magnetic properties locally, electric-field control of domain walls in PMA systems is also pursued. In previous studies, electric fields are either applied across a gate dielectric,\cite{SCH-12,CHI-12,BER-13,FRA-13} an ionic conductor,\cite{BAU-13,BAU-15} or a piezoelectric transducer.\cite{SHE-15} These approaches manipulate the strength of PMA. As a result, the velocity or pinning of magnetic domain walls that are driven by a magnetic field or electrical current is deliberately tuned. Since the underlying physics relies on the variation of PMA, electric field control is most pronounced for thermally activated creep motion.

ln this letter, we report on a PMA system wherein domain walls are actively driven by an electric field. Our approach differs from previous work in that the electric field does not modify the strength of PMA. Instead, electric fields are used to move domain walls in a ferroelectric crystal onto which magnetic domain walls in a neighboring PMA multilayer are strongly pinned. The magnetic domain walls are thus driven by an electric field without the assistance of a magnetic field or electrical current. Reversible motion of domain walls and an exponential increase of domain wall velocity with electric field strength are demonstrated.       

\begin{figure}[t]
\includegraphics{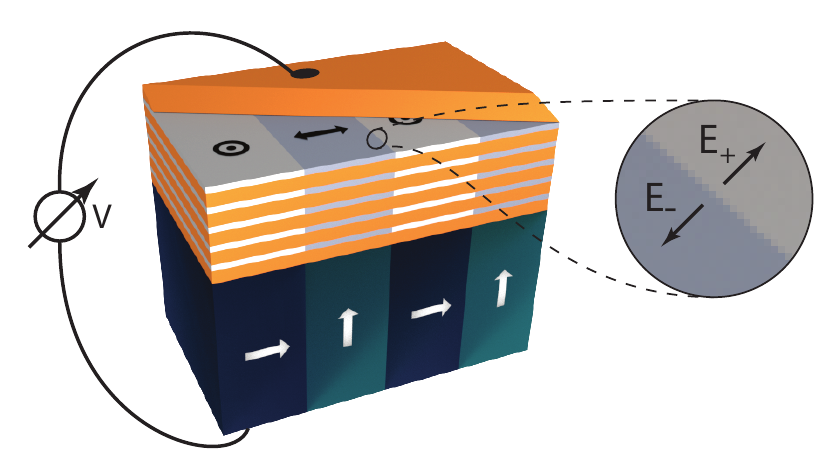}
\caption{\label{Fig1} Illustration of domain correlations in our experimental sample. The ferroelectric polarization of the BaTiO$_3$ substrate alternates between in-plane and perpendicular directions (white arrows). Strain transfer from these domains to the [Cu (9 nm)/Ni (2 nm)]$_{5}$/Cu (9 nm) multilayer causes the magnetization to rotate from perpendicular to in-plane (black marks). The magnetic domain walls are strongly pinned onto the domain walls in the ferroelectric substrate, enabling reversible motion in positive and negative out-of-plane electric fields.}
\end{figure}

\begin{figure}[t!]
\includegraphics{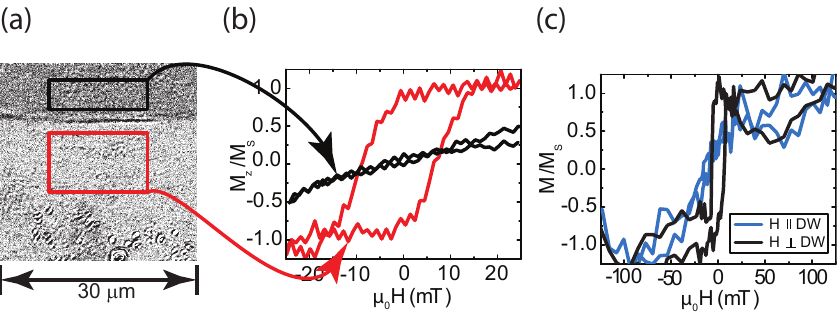}
\caption{\label{Fig2} (a) MOKE microscopy image of the Cu/Ni multilayer. (b) Polar MOKE hysteresis curves for two neighboring domains. The rectangles in (a) illustrate the area of data collection. The measurements indicate that one of the domains exhibits PMA (domain with red rectangle), while the magnetization of the other domain is oriented in-plane (domain with black rectangle). (c) Local longitudinal MOKE hysteresis loops for the domain with in-plane magnetic anisotropy, i.e., both curves are measured in the black rectangle. The orientation of in-plane magnetic field is parallel (blue curve) and perpendicular (black curve) to the domain wall. The magnetic domain wall does not move in a magnetic field because of strong elastic pinning to a ferroelectric domain wall in the BaTiO$_3$ substrate.}
\end{figure}

We consider a [Cu (9 nm)/Ni (2 nm)]$_{5}$/Cu (9 nm) multilayer on a single-crystal ferroelectric BaTiO$_3$ substrate. The Cu and Ni layers are grown by molecular beam epitaxy at room temperature onto a 1 nm Fe buffer layer and capped by 5 nm Au. Before sample preparation, the polarization of the BaTiO$_3$ substrate is oriented in-plane. Despite a lattice mismatch of more than 10\% between the metal layers and BaTiO$_3$, the Cu/Ni multilayer grows epitaxially with a (001) orientation. Instant strain relaxation during MBE, decouples the Cu/Ni multilayer from the BaTiO$_3$ substrate. As a result, the Ni layers are strained only by the lattice mismatch with Cu. The growth-induced tensile strain in Ni amounts to 1.9\%, which is sufficient to induce PMA via a strong magnetoelastic effect.\cite{JOH-96,PER-08} For more details on the crystallographic structure of the Cu/Ni multilayer on BaTiO$_3$, we refer to Ref.\citenum{SHI-15}, which reports on magnetization switching. Here, we focus on the dynamics of electric-field-driven domain wall motion.   

Domain walls are created in the Cu/Ni multilayer by the application of a strong out-of-plane electric field across the ferroelectric BaTiO$_3$ substrate. After the field is turned off, the BaTiO$_3$ crystal relaxes into stripe domains with alternating perpendicular and in-plane polarization. Since the strain in Ni layers on top of ferroelectric domains with in-plane polarization does not change by this procedure, PMA is preserved in these areas. On top of domains with perpendicular polarization, this is not the case. Polarization switching from in-plane to perpendicular coincides with a 90$^\circ$ rotation of the tetragonal BaTiO$_3$ unit cell. In the substrate plane, this corresponds to a uniaxial lattice compression of 1.1\%.\cite{LAH-12} Now that the multilayer is clamped to the substrate, voltage-induced strain effects are efficiently transferred to the Cu/Ni multilayer. The growth-induced tensile strain in Ni layers on domains with perpendicular polarization is thus considerably reduced and this causes the magnetization to rotate into the film plane. Figure \ref{Fig1} illustrates the ferromagnetic and ferroelectric domain configuration after the application of a strong out-of-plane electric field.  

We use magneto-optical Kerr effect (MOKE) microscopy to characterize the sample. In Fig. \ref{Fig2}, we focus on an area of the Cu/Ni multilayer with one magnetic domain wall. Local polar MOKE microscopy data (Fig. \ref{Fig2}(b)) confirm that one of the domains exhibits PMA (bright domain in Fig \ref{Fig2}(a)). Using vibrating sample magnetometry with large in-plane magnetic field (not shown), we derive an effective PMA of $K_\mathrm{eff}=2.4\times10^4$ J/m$^3$ for this domain type. The polar MOKE hysteresis curve of the other domain is completely closed (black curve in Fig. \ref{Fig2}(b)). Local longitudinal MOKE microscopy measurements on this domain (Fig. \ref{Fig2}(c)) indicate that the in-plane magnetization aligns perpendicular to the domain wall. From the slope of the hard-axis hysteresis curve, we estimate an uniaxial in-plane magnetic anisotropy of $K_\mathrm{u}=1.2\times10^4$ J/m$^3$. The magnetic domain wall in the Cu/Ni multilayer is pinned onto a ferroelectric domain wall in the BaTiO$_3$ substrate, separating domains with in-plane and perpendicular polarization. This strong pinning effect, which prevents any domain wall displacement in an applied magnetic field, is caused by the abrupt change in magnetoelastic anisotropy.       

\begin{figure}[t!]
\includegraphics{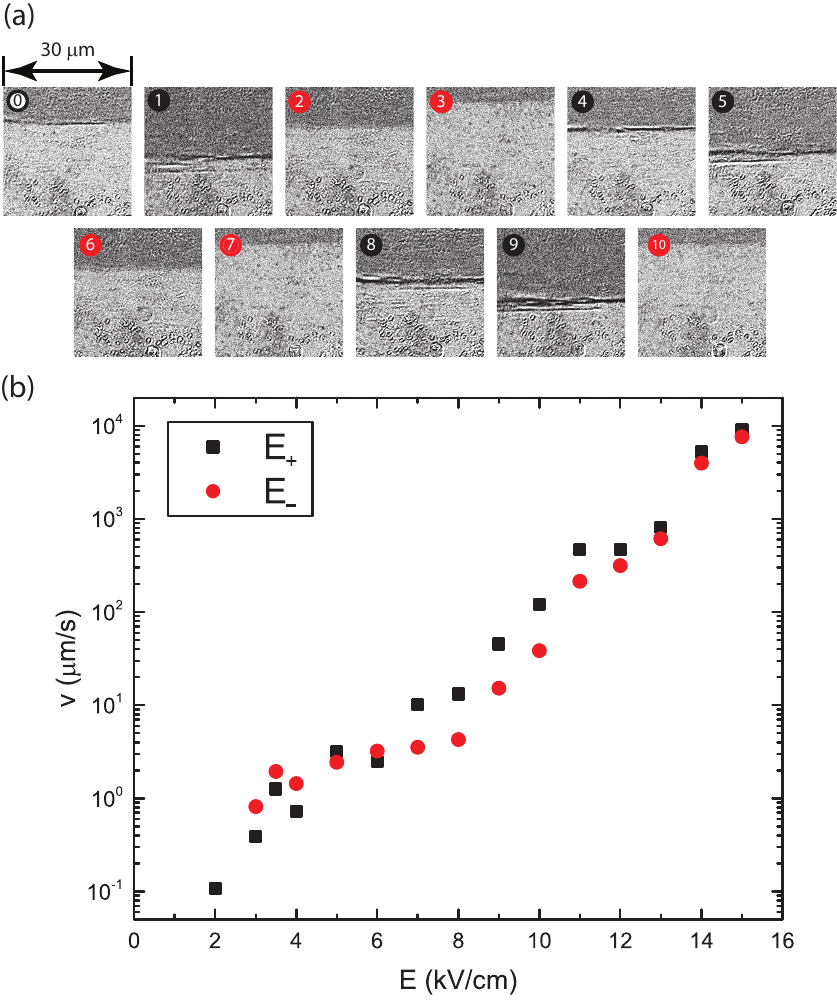}
\caption{\label{Fig3} (a) Sequence of MOKE microscopy images demonstrating reversible electric-field-driven magnetic domain wall motion in the Cu/Ni multilayer. Images with black (red) dots are recorded after the application of a positive (negative) field pulse across the BaTiO$_3$ substrate. The strength of the electric field is $\pm$3.5 kV/cm. (b) Dependence of domain wall velocity on electric field strength.}
\end{figure}

Next, we discuss the magnetic response to out-of-plane electric field pulses. In the experiments, the Cu/Ni multilayer is grounded and square voltage pulses are applied to a metallic electrode on the back of the BaTiO$_3$ substrate. Positive pulses generate an electric field ($E_\mathrm{+}$) favoring ferroelectric domains with polarization pointing upward. As a result, these domains grow at the expense of neighboring domains with in-plane polarization. A reversal of the voltage polarity (i.e. negative voltages or $E_\mathrm{-}$) causes ferroelectric domains with upward polarization to shrink. Alternating positive and negative voltage pulses thus reversibly move the ferroelectric domain walls in the BaTiO$_3$ substrate. Figure \ref{Fig3}(a) shows the magnetic response of the Cu/Ni multilayer to this electric field effect. Positive electric field pulses (indicated by black numbered dots) displace the magnetic domain wall downward. In this case, the domain with in-plane magnetization grows at the expense of the magnetic domain with PMA. Negative out-of-plane electric field pulses (red numbered dots), in turn, move the magnetic domain wall back up. Reversible motion of the domain wall is caused by strong strain coupling between magnetic domains in the Cu/Ni multilayer and ferroelectric domains in the BaTiO$_3$ substrate. Correlations between the different ferroic domains are preserved in an applied electric field, as previously demonstrated for magnetic films with in-plane magnetic anisotropy.\cite{LAH-11} Consequently, the magnetic and ferroelectric domain walls move in unison. Electric-field-driven motion of the magnetic domain wall does not require any assistance from a magnetic field or electrical current.         

\begin{figure*}[t!]
\includegraphics{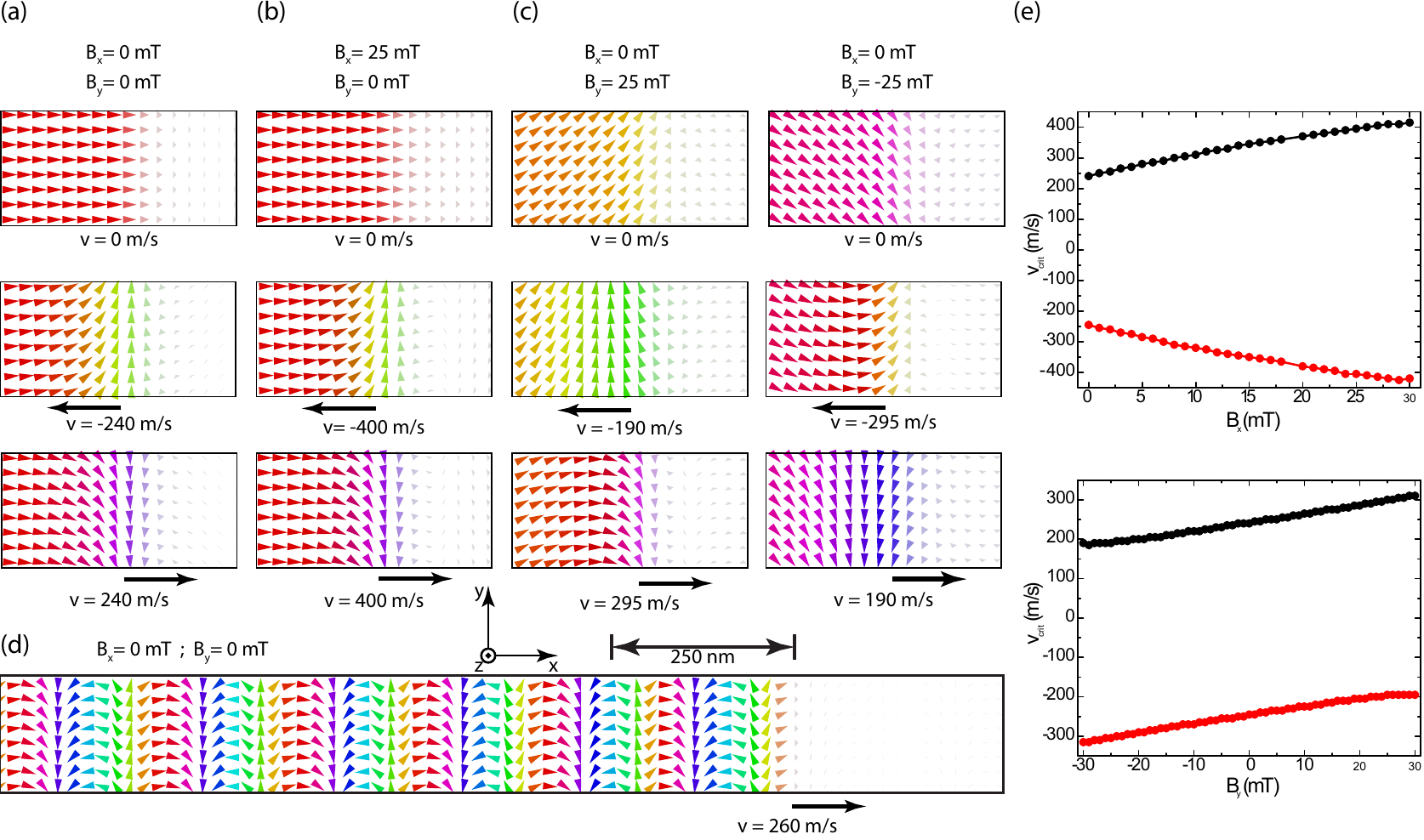}
\caption{\label{Fig4} (a)-(c) Micromagnetic simulations of the magnetic domain wall for zero and non-zero magnetic fields along the $x$- and $y$-axis. Colored arrows indicate the in-plane orientation of magnetization. The magnetization is oriented out-of-plane in white areas. In (a)-(c), $v<v_\mathrm{crit}$. (d) Spin wave emission from a moving domain wall for $v>v_\mathrm{crit}$. The double-headed arrow indicates the scale of all micromagnetic simulations. (e) Variation of critical domain wall velocity with magnetic bias field along the $x$- and $y$-axis.}
\end{figure*}

The domain wall velocity ($v$) varies with the strength of applied electric field. To extract the $v-E$ dependence, we adjust the duration of the voltage pulses ($\Delta$$t$) to the strength of the applied electric field, so that the displacement of domain walls is accurately measured (typically $\Delta$$x$ $\approx$ 5 $\mu$m). From the data, the velocity ($v$) of electric-field-driven magnetic domain walls is derived ($v=\Delta$$x$/$\Delta$$t$). Figure \ref{Fig3}(b) summarizes the results for positive and negative out-of-plane field pulses. For both pulse polarities, the domain wall velocity increases by five orders of magnitude when the electric field is enhanced from 2 kV/cm to 15 kV/cm. The exponential variation of $v$ with $E$ is explained by thermally assisted depinning of the ferroelectric domain wall in the BaTiO$_3$ substrate.\cite{MIT-53} The maximum domain wall velocity in our proof-of-concept experiments is 1 cm/s. Despite being relatively small compared to current-driven magnetic domain wall motion,\cite{MIR-11,EMO-13,RYU-13} it is three orders of magnitude faster than our previous result on 20-nm-thick Fe films with in-plane magnetic anisotropy.\cite{FRA-15} This observation demonstrates that exponential scaling of the domain wall velocity persists up to large electric field. Further enhancements of the maximum velocity would require the use of ultrashort and even stronger field pulses. An extrapolation of the data in Fig. \ref{Fig3}(b) suggests that $E\approx$ 30 kV/cm could possibly increase the domain wall velocity to 100 m/s. For reasonably small voltages, this is only achieved when the thickness of the BaTiO$_3$ layer is reduced from 500 $\mu$m (this study) to $<$1 $\mu$m. Since it is not obvious how downscaling affects the dynamics of ferroelectric domain walls, this open question needs to be addressed in future studies.     

To assess domain wall dynamics at high velocities, we performed micromagnetic simulations using MuMax3.\cite{VAN-14} In the simulations, a 2-nm-thick Ni layer is discretized into $2.5\times2.5$ nm cells. As input parameters, we use the magnetic anisotropy values from our experiments, together with a saturation magnetization $M_\mathrm{s}=4.8\times10^5$ A/m,\cite{KIT-49} an exchange constant $A=7.3\times10^{-12}$ J/m,\cite{LAS-01} and a damping parameter $\alpha$ = 0.015. The application of an out-of-plane electric field is mimicked by fast lateral motion of an anisotropy boundary that separates domains with perpendicular and in-plane magnetization. Domain wall motion at velocity $v$ is implemented by shifting the anisotropy boundary over one discretization cell ($\Delta$$x$ = 2.5 nm) during each time window $\Delta$$t=\Delta$$x$/$v$ of the simulation. Two-dimensional periodic boundary conditions are used. The micromagnetic simulations thus reflect the dynamic response of a continuous Ni/Cu multilayer on top of BaTiO$_3$.

Figure \ref{Fig4} shows an overview of simulation results. Under static conditions and zero magnetic field (Fig. \ref{Fig4}(a)), the simulations reproduce the domain configuration in the MOKE microscopy images of Figs. \ref{Fig2} and \ref{Fig3}, i.e., one domain with in-plane magnetization pointing towards the domain wall and another domain with PMA. The 90$^\circ$ magnetic domain wall that separates the two domains is of the N\'{e}el type. At high domain wall velocities $v$, the magnetization within the wall tilts towards the $y$-axis. The tilt angle depends on the direction of domain wall motion and steadily grows with $v$. Above a critical velocity $v_\mathrm{crit}$, here corresponding to the velocity at which the magnetization points along $+y$ or $-y$, the spins within the magnetic domain wall start to precess continuously. This dynamic behavior, which is similar to Walker breakdown in magnetic domain walls that are driven by a magnetic field or electric current,\cite{SCH-74, THI-05,SHI-11} results in the emission of spin waves.\cite{VAN-14b} Figure \ref{Fig4}(d) shows an example.  

The application of a magnetic field does not move the magnetic domain wall, in agreement with experiments, but it changes its spin structure (Fig. \ref{Fig4}(c)). Besides, a magnetic field either stabilizes or destabilizes the domain wall at high velocities, depending on the directions of applied magnetic field and domain wall motion. In general, external magnetic fields that suppress magnetization tilting towards the $y$-axis enhance the critical velocity. The simulated variations of $v_\mathrm{crit}$ with in-plane magnetic fields along the $x$- and $y$-axes are summarized in Fig. \ref{Fig4}(e).

In summary, we have demonstrated reversible electric-field-driven motion of magnetic domain walls in perpendicularly magnetized Cu/Ni multilayers on ferroelectric BaTiO$_3$ substrates. The driving mechanism is based on strong elastic coupling between magnetic and ferroelectric domains and does not require assistance from a magnetic field or electrical current. The velocity of the magnetic domain walls is fully determined by the electric field strength. While not affecting magnetic domain wall motion, magnetic fields allow for versatile tuning of their internal spin structure. Electric and magnetic fields can thus be used to independently move and tailor magnetic domain walls in Cu/Ni multilayers. \\ 

This work was supported by the European Research Council (ERC-2012-StG 307502-E-CONTROL and ERC-2014-PoC 665215-EMOTION), JSPS KAKENHI (Grant Nos. 15H01998, 15H01014, 16K14381), the Creation of Life Innovation Materials for Interdisciplinary and International Researcher Development Project of MEXT, and the Collaborative Research Project of the Materials Structures Laboratory, Tokyo Institute of Technology.


%


\end{document}